\documentclass[journal]{vgtc}              

\onlineid{0}

\vgtccategory{Research}

\vgtcpapertype{please specify}

\renewenvironment{quote}{%
  \list{}{%
    \leftmargin0.25cm   
    \rightmargin\leftmargin
  }
  \item\relax
}
{\endlist}

\def\BibTeX{{\rm B\kern-.05em{\sc i\kern-.025em b}\kern-.08em
    T\kern-.1667em\lower.7ex\hbox{E}\kern-.125emX}}

\title{Locatability and Locatability Robustness of Visual Variables in Single Target Localization}

\author{%
Wei Wei, Miguel A. Nacenta, Michelle F. Miranda, and Charles Perin
}

\authorfooter{
  \item Wei Wei, Miguel A. Nacenta, Michelle F. Miranda and Charles Perin are with the University of Victoria. E-mail: weiwei@uvic.ca, nacenta@uvic.ca, michellemiranda@uvic.ca, cperin@uvic.ca.
}

\abstract{
Finding a particular object in a display is important for viewers in many visualizations, for example, when reacting to brushing or to a highlighted object. This can be enabled by making the target object different in one of the visual variables that determine the object's appearance; for example, by changing its color or size. Certain interpretations of the visual search literature have promoted the view that using visual variables such as hue---often labeled as \textit{preattentive}---would make the target object automatically ``popout,'' implying that an object can be located almost instantly, regardless of the number of objects in the display. In this paper we present a study that serves as a bridge between the extensive visual search literature and visualization, establishing empirical base measurements for the localization task. By testing displays with up to hundreds of objects, we are able to show that none of the common visual variables is immune to the increase in the number of objects. We also provide the first empirically informed comparisons between visual variables for this task in the context of visualization, and show how different visual variables have varying robustness with respect to two additional dimensions: the location of the target and the overall visual arrangement (layout). A free copy of this paper and all supplemental materials are available
on our online repository: \url{https://osf.io/z68ak/overview}.
}

\keywords{Visual variables, visual search, selectivity, popout effect, preattentiveness, visualization.}

\usepackage[final]{changes}
\usepackage{mathptmx}                  
\usepackage{algorithmic}
\usepackage{array}
\usepackage[caption=false,font=normalsize,labelfont=sf,textfont=sf]{subfig}
\usepackage{textcomp}
\usepackage{stfloats}
\usepackage{url}
\usepackage{verbatim}
\usepackage{graphicx}
\usepackage{balance}
\usepackage{amsmath,amssymb}
\usepackage{enumitem}
\usepackage{xcolor}
 
\usepackage{bm}
\usepackage{tabularx}
\usepackage{multirow}
\usepackage{colortbl}
\usepackage{rotating}
\usepackage{bigstrut}
\usepackage{makecell}
\usepackage{xspace}

\newcommand{\HLlayout}{$\bm H_{layout}$\xspace}
\newcommand{\HLlocation}{$\bm H_{location}$\xspace}
\newcommand{\HLsetSize}{$\bm H_{set size}$\xspace}
\newcommand{\HLvariable}{$\bm H_{variable}$\xspace}
\newcommand{\HRsetSizegrid}{$\bm H_{R:set size,GRID}$\xspace}
\newcommand{\HRsetSizeNongrid}{$\bm H_{R:set size,NON-GRID}$\xspace}
\newcommand{\HRdisToCengrid}{$\bm H_{R:Dis\_TarCen,GRID}$\xspace}
\newcommand{\HRdisToCenNongrid}{$\bm H_{R:Dis\_TarCen,NON-GRID}$\xspace}
\newcommand{\inlinesection}[1]{\vspace{0.25em}\noindent$\blacktriangleright$~\textsc{\textbf{#1}}\xspace}
\newif\ifcoloron
\coloronfalse  

\newcommand{\changeColor}[1]{%
  \ifcoloron
    \textcolor{blue}{#1}%
  \else
    #1%
  \fi
}

\graphicspath{{figs/}{figures/}{pictures/}{images/}{./}} 

\begin{document}

\firstsection{Introduction}
\label{sec:introduction}
\maketitle
Locating a specific object among many others is a fundamental task in information visualization (e.g., pointing out which dot in a scatterplot corresponds to a given data point or focusing attention on a highlighted data point). It is also a subtask of higher-level visualization tasks, such as visual comparison since one must first locate the visual objects before comparing them.  
One common way to support this task is to create differences in a visual variable of the target object, for example, by changing its color~\cite{Griffin2015} or size~\cite{Li2010}.

\changeColor{
In contrast with the many recommendations around the use of visual variables for the task of magnitude comparison (e.g., the lists of visual variables ordered by accuracy provided in textbooks~\cite[p.102]{Munzner_2014} \cite[p.168]{Ware_2012}and study results~\cite{Cleveland_1984,Heer2010}), we lack the empirical foundation to allow visualization designers and researchers to make informed choices for object localization. We believe that this is due to a combination of factors. 
First, while the visual search literature is extensive, it is often inconsistent in terms of terminology (e.g., different authors use terms such as visual search efficiency~\cite{Frintrop2010,Wolfe1998Visual, wolfe2017five}, preattentiveness~\cite{Julesz_1983,Treisman1985Preattenive} and popout~\cite{Egeth1972, Treisman1985} to refer to properties of visuals in different tasks), it also does not directly apply to visualization (e.g., the number of objects tested in visual search studies is often small compared to the number of objects in many visualizations); and the results are not definitive at a meta-level. 
Second, although the appropriateness of visual variables for target localization was explored early in visualization community (Bertin calls some visual variables ``selective''~\cite[p. 48]{bertin1983semiology}), the little attention devoted to the localization task has focused on whether variables are preattentive or not (e.g., Ware notes: ``If processing is preattentive, the time taken to find the target should be independent of the number of distractors''~\cite[p. 149]{Ware_2012}). 
However, recent visual search literature has shown that the time it takes to find a target generally increases as the total number of objects increases~\cite{Wolfe2021}, so visual variables are not simply preattentive or not~\cite{Wolfe1998Visual,Wolfe_2004,Wolfe2021}.
}

The consequences of the current state of understanding might be costly: we tend to ignore a fundamental task in information visualization, and we cannot effectively incorporate existing empirical knowledge in designs or new research. 
As a first step to address this problem, we carried out two experiments ($N=2\times 112$) that compared seven visual variables (\textsc{Luminance}, \textsc{Single-Hue}, \textsc{Hue}, \textsc{Size}, \textsc{Orientation}, \textsc{Length}, and \textsc{Shape}) in terms of \emph{locatability} and \emph{locatability robustness}. We defined these two terms to circumvent existing terminological issues in the visual search and visualization literature. The experiments consider set size (i.e., number of objects) and two more dimensions: target location (\textsc{Center} vs.\ \textsc{Periphery}) and layout (\textsc{Grid} vs.\ \textsc{Non-grid}). The results provide the first comparisons between visual variables in terms of locatability (the time it takes to detect and gaze-fixate the target), and the robustness of each visual variable with respect to the changes of set size, target location, and object layout.

Beyond the increased understanding of the target localization task, we provide a queryable model that provides evidence-based support for design and research, as well as a careful review of previous research that can help build a bridge between existing research in visual search and its potential application in the domain of visualization.

\section{Background and Related Work}
\label{sec:background}
We review prior research in cognitive science on visual search and the concept of preattentiveness, along with visualization studies concerned with practical applications of preattentiveness. 
Our review suggests that some aspects of the concept, particularly its conceptualization and quantification, remain open to further investigation.

\subsection{Visual Search and Preattentive Features}
\label{sec:visualsearch}


Humans engage in the cognitive process of visual search daily, often without realizing it.
For example, we routinely locate our phones, browse for interesting news on them, and scan crowds to find friends. 
Psychophysicists have conducted many experiments to investigate the mechanisms of visual search. 
Typically, participants are asked to answer a question as follows: \textit{given a target and a test image, is there an instance of the target in the test image?}~\cite{Frintrop2010} 
Most experiments manipulate two independent variables: set size---the number of objects on display (targets and distractors), and visual features---the attributes distinguishing targets from distractors (e.g., color if participants must identify a red object among blue objects).
See supplementary materials for a table of representative studies.

Performance in visual search is measured by participants' reaction time (RT), that is, the time between stimulus presentation and the participant’s response indicating target detection.
RT is then used to quantitatively examine which visual features facilitate shorter RT and are resilient to changes in set size.  
Researchers often model RT as a linear function of set size and examine its slope~\cite{Francolini1979,Treisman1985,Treisman_1988} (which has been called ``search efficiency''~\cite{Frintrop2010,Wolfe1998Visual, wolfe2017five} or ``search rate.''~\cite{Francolini1979,Treisman1985,Treisman_1988}) A less steep slope indicates greater resilience of a visual feature to increasing set sizes. 
Studies have identified a set of features with \textit{nearly} zero slope, meaning RT remains almost constant as set size increases. 
This phenomenon, informally termed popout~\cite{Egeth1972, Treisman1985}, is attributed to preattentive processing~\cite{Treisman1985Preattenive}, wherein target detection occurs rapidly (within milliseconds) and effortlessly.

The study of preattentive features in psychology has a long history (see~\cite{TheHandbookofAttention,Wolfe2014Approaches} for reviews). 
Notably, the Feature Integration Theory (FIT) models the visual search with a two-stage process~\cite{Treisman_1980}: (i) a preattentive stage, where basic features such as color are processed in parallel across the visual field, then (ii) an attentive stage, where focused attention serially binds these features into coherent perceptual objects. 
The features that have been identified as preattentive include
color~\cite{Farmer1980,Carter1982,DZmura1991, Nagy1990, Bauer1998,Brawn1999,Healey1999,Daoutis2006}, 
size (including length)~\cite{Cavanagh1990,Found1996,Stuart1993, Treisman_1988}, 
orientation~\cite{Moraglia1989,Foster1991Asymmetries, Bergen1983,Cavanagh1990},
and motion~\cite{Dick1987,Braddick1991,McLeod1988}.
Some other features, like shape, are found to not be \textit{truly} preattentive~\cite{Chen1982, Kristjnsson2001, Giovannangeli_2022}. 


\subsection{Preattentiveness in Visualization Studies} %
\label{sec:related_work}

Visualization researchers routinely apply the concept of preattentiveness to visualization design~\cite{Barrera-Leon_2022,Krekhov2019,Gramazio_2014,Boger2021,Elliott2021}.  
\added{
Preattentiveness is vaguely considered a property of a visual variable, or a specific mechanism in which a visual variable is perceptually processed. Preattentiveness makes the variable ``popout'' or ``efficient,'' and therefore makes certain variables better design choices. 
The concept of \emph{visual variable} (introduced by Bertin~\cite{bertin1983semiology}, also referred to as ``visual channel'' by Munzner~\cite{Munzner_2014} and ``visual feature'' by Ware~\cite{Ware_2012} ) is largely accepted as a key consideration for visualization design; it appears in major textbooks and anthologies (e.g.,~\cite{bertin1983semiology,Ware_2012,Munzner_2014,Spence2014}). 
However, these texts often recommend variables based on how accurate humans are at estimating or comparing variations in visual variables. 
Preattentiveness, popout or search efficiency are often only mentioned without empirical support, perhaps due to the issues that we detail in the next subsection.}

In this paper, we focus on static 2D variables, although dynamic variables like motion have been studied in 
visualization as well (e.g.,~\cite{Huber2005, Heer2007, Chevalier2016,Waldner2014,Gutwin_2017,Chalbi2019,Bartram2002}). 
Color is the variable that is most studied in terms of preattentiveness. Researchers have investigated both detection and estimation (e.g., value comparisons) of color. 
For example, in a series of studies, Healey et al. found that hue can be useful for boundary detection~\cite{Healey_1995} and numerical estimation~\cite{Healey_1996} in data analysis.
Color has been extensively studied to encode nominal information~\cite{Bianco2014, moreland2009diverging,silva2011using, lin2013selecting}. Here, discriminability between colors is an important factor to consider.
Healey proposed to maximize both the total number of colors and the perceptual distances between colors in color palettes to make colors more discriminable from each another~\cite{Healey96Choosing}. 
Follow-up research (e.g., Maxwell et al.~\cite{Maxwell2000}, ColorBrewer~\cite{Harrower2003}, Colorgorical~\cite{Gramazio2017}, color name models~\cite{heer2012color} and color difference models~\cite{Szafir2018}) looked further into creating discriminable and aesthetic color palettes for visualization.
Color is also useful to highlight information in dashboards~\cite{Yalcin2018}, in coordinated multiple views~\cite{Griffin2015}, and to show visual links between visualizations~\cite{Steinberger2011}.

Length and size are less utilized than color for their preattentive capabilities in visualization, 
since variations in length and size tend to be used to encode quantitative data attributes~\cite[p. 102]{Munzner_2014}---for example, length in bar charts and size (or area) in bubble charts or treemaps. 
However, using these two variables for detection rather than estimation is useful in some visualizations---for example, to highlight a data point in a scatterplot by making it larger than the others. 
Size has been shown to provide a more dominant visual contrast than shape in scatterplots~\cite{Li2009}, and models have been proposed to help designers pick optimal symbols' size for discriminability in scatterplots for analytic tasks~\cite{Li2010}. 
Length and size are also used to make graphical elements popout beyond charts. 
For example, in font design, font attributes like bold, squished, and condensed leverage length and size to enhance text emphasis~\cite{Brath2016,lang_infotypography}, and increasing font size has been found to be the best technique for text highlighting~\cite{Strobelt2016}. 

The preattentiveness of variables such as orientation is less explored in visualization, as noted in previous research~\cite{Hall2016}. 
To our knowledge, only Healey et al.~\cite{Healey_1996} have tested orientation. 



\subsection{Conceptual Precision and Quantification}
\label{background:gap}

\changeColor{
While the literature on visual search is extensive, relevant concepts and measurements remain challenging to operationalize, particularly when applied to visualization. 
In this section, we discuss the terminological challenges that led us to adopt different terms for this study, which we introduce and justify in Section~\ref{sec:locatability}.
}

\added{
A first issue is that the preattentiveness, popout or search efficiency of a visual variable are often described as intrinsic characteristics of the visual variable, despite the fact that these terms often refer to related---but ultimately very different---tasks. Specifically, the literature on the topic considers three main types of tasks. \emph{Detection} tasks require \textit{indicating the presence or absence of a target} (e.g.,~\cite{Egeth1984,Treisman1985, Treisman_1988, DZmura1991,Palmer1993}). 
\emph{Identification } tasks require \textit{classifying the target into one of several types} (usually two)~\cite{Treisman_1980,Baldassi2002}. 
\emph{Localization} tasks require \textit{reporting the target's location on the display~\cite{Hyun2009,Atkinson1989}.} 
We have observed that existing literature tends to inadvertently assume that preattentiveness or popout are intrinsic characteristics of visual variables rather than properties of how humans perform with them in a particular task (e.g.,~\cite{Healey_2012,griouicomparing}).
We think these assumptions lead to misconception and overgeneralization of existing empirical results. Therefore, we focus on localization and make it clear that our empirical results only apply to the localization task by including the task name in the nomenclature (Section~\ref{sec:locatability}). 
We chose localization because it is (i) integral to many low-level activities in visualization and visual analytics, such as filtering data points and finding anomalies~\cite{Amar2005}; and (ii) as we will discuss in Section~\ref{sec:diss:term_and_tasks}, neither too low-level nor too high-level. 
}

A second issue is that previous literature, both in visual search and in visualization, often conceptualizes the preattentiveness or efficiency of a visual variable in a binary way, based on arbitrary thresholds. 
For example, Ware~\cite{Ware_2012} suggests a threshold for search efficiency at less than 10 ms per item; Wolfe a threshold ``near (but typically a little greater than) 0 ms/item,''~\cite{Wolfe2021}
and others propose fixed thresholds of 100 ms~\cite{Moreira_2020}, 200-250 ms~\cite{Healey_2012}, and 500 ms~\cite{VanderPlas2019} regardless of set size. 
However, evolving perspectives in the visual search domain reject these binary terms and instead define these properties on a continuum (e.g.,~\cite{Wolfe1998Visual,Wolfe_2004,Wolfe2021}). 
Moreover, our review of previous studies shows that experiments rarely test more than 40
objects (with few exceptions, still below 60 objects~\cite{Nagy1990}), in stark contrast with many visualizations, which can contain hundreds of objects or more. 
The lack of evidence with larger, visualization-realistic numbers, and the dichotomous view of preattentiveness might lead us to ignore important differences in time costs with different variables.

Similarly to large set sizes, there are factors that are of practical relevance to visualization, but not to the psychophysics community. 
This is probably because features such as target location and spatial layout do not necessarily offer perceptual insight, yet might affect performance in visualization. 
We know, for example, that central versus peripheral positioning affects object identification~\cite{Gutwin_2017}, and that layout can modulate attentional capacity~\cite{Haroz2012}. 
In our research we wanted to consider the effects of these factors and we wondered whether they are independent or related to the effect of set size.

\section{Locatability and Locatability Robustness}
\label{sec:locatability}

\changeColor{
As we have argued in Section~\ref{background:gap}, existing terminologies from the literature pose a challenge to us because they might refer to different tasks or to slightly different constructs, are artificially dichotomous, and do not consider factors that might be important in visualization as an application domain. To circumvent those challenges, we define two new terms: \emph{locatability} and \emph{locatability robustness}. These do not reflect the discovery of new phenomena; instead, they are a relabeling and realignment of existing constructs that facilitate the applicability of this research. We define locatability as follows:
}

\begin{quote}
    $\blacktriangleright$ \emph{The \textbf{locatability} of a visual variable for a defined localization task is the performance with which a human can visually locate a target among a set of distractors, with the target having a different level on that visual variable than the distractors.}
\end{quote}
\changeColor{We measure performance in terms of task completion time because (i)~we think this metric has the most significance for the task (near-perfect accuracy is confirmed by our results) and (ii)~time has been the dominant metric used in previous studies of this and similar tasks.
A visual variable is more locatable if it takes less time to locate targets with that variable.}
We also define locatability robustness as follows:
\begin{quote}
    $\blacktriangleright$ \emph{The \textbf{locatability robustness} of a visual variable with respect to a factor is the degree to which locatability changes as that factor changes.}
\end{quote}
The more robust a visual variable, the less its locatability changes with changes in the factor.
For example, hue might be more robust with respect to set size than shape if the locatability of hue changes less with variations of the set size than the locatability of shape. 
Robustness can also be affected by qualitative factors, such as the type of layout.

How do these terms differ from, or relate to, existing ones? 
\added{
Locatability is similar to popout when the latter is used in the context of localization tasks, but is different in that it does not presuppose a particular outcome, and does not suggest dichotomy (as in ``visual variable X is popout or not''). 
In contrast to preattentiveness, the notions of locatability and locatability robustness do not presuppose any specific underlying perceptual or cognitive mechanism. Additionally, our terms are defined explicitly for localization, and do not encompass multiple possible tasks (such as detection and identification). This should reduce the chance that researchers and practitioners use or interpret those terms without considering which task they refer to.
Moreover, the term locatability robustness is also more versatile and general in scope than search efficiency or search rate because it characterizes how performance in localization tasks varies not only with respect to set size, but also to other factors such as target location and type of layout.}

\section{Visual Variables and Factors of Interest}
\label{sec:factorsofinterest}

To characterize visual variables in terms of their locatability and locatability robustness, it is necessary to scope the factors and conditions to study.
We selected visual variables and factors with practical meaning for visualization design.

\textbf{In terms of visual variables}, \added{we compiled a list of variables that: (i) are discussed in textbooks, and (ii) are practical and commonly used in visualization design. Three authors convened and discussed each variable, aiming to strike a balance between completeness and practicality (including all variables would require too many trials or compromise statistical power). 
The final list includes}
\textsc{Luminance}, \textsc{Single-Hue}, \textsc{Hue}, \textsc{Size}, \textsc{Orientation}, \textsc{Length}, and \textsc{Shape}. 
We included the first three variables to examine the potential nuances among commonly differentiated aspects of color~\cite{Stone2016}. 
They also represent different usages of color in visualization: luminance and single-hue (a simultaneous variation of saturation and luminance popular in practice) are generally used to encode ordinal values, while hue is recommended for nominal values~\cite[p. 102]{Munzner_2014}.
We selected size, orientation, length, and shape that are generally accepted as part of the canon of existing variables (e.g.,~\cite{bertin1983semiology, Cleveland_1984, Mackinlay_1986,quadriSurveyPerception2022,Munzner_2014,Heer2010}). 
Although length, size, and orientation are established as preattentive, shape is categorized as probably preattentive~\cite{Wolfe_2004}. Yet, we included shape because it is 
often used to encode nominal values in scatterplots~\cite{Lewandowsky1989,Smart2019}. 

\textbf{In terms of factors}, we included set size (the number of objects on screen), 
layout, and target location.

For set size, we selected values (12, 48, 192 and 768) that cover a wide range of object density: from almost trivial (locating an object among very few) to 
hundreds of objects (much larger than the typical set sizes tested in similar studies). 
Values increase exponentially so that we could examine the time-set size relationship on a logarithmic scale.

Layout refers to the regularity of the arrangement of objects on the display. 
We selected two categories of layouts: \textsc{Grid} and \textsc{Non-grid}. 
Grid layouts are representative of regularly arranged visualizations such as 2D bar charts, heatmaps~\cite{Manteau2017Reading,hanEffectVisualInteractive2020}, tables with in-cell visual encodings~\cite{jieffect,tablelens,bertifier}, and small multiples~\cite{tufte1983visual}). 
Typical visualizations with non-grid layouts are those where marks are not necessarily aligned (e.g., treemaps~\cite{Kong2010}, arc diagrams~\cite{Wattenberg2002} and node-link diagrams~\cite{Saket2014}), or the visual variable position is determined by a data attribute (e.g., scatterplots~\cite{Micallef2017,Sarikaya2018} and maps~\cite{Wood2011}).


Target location refers to where the target is located on the display. 
This factor is often beyond the control of the designer because position on the plane is considered the most effective visual variable to encode quantitative values (e.g.,~\cite{bertin1983semiology, Cleveland_1984, Heer2010, Munzner_2014}) and therefore the position of objects is often data-driven.
Yet, we included location because it can help us explain advantages of certain visualization techniques. Indeed, the ability to perceive the variation of a visual variable peripherally---before directly focusing on it---is likely to influence how fast one can identify targets~\cite{Gutwin_2017}. 
We manipulate whether the target is in \textsc{Center} or in \textsc{Periphery} of the display---a nominal factor that we refer to as target location---and consider how the distance between the target and the center point of the display affects performance, a continuous measurement that we refer to as Dis\_TarCen.

\section{Hypotheses}\label{research-questions}
We address two research questions that translate to two sets of hypotheses: 
(a) what is the effect on locatability of each factor (visual variable --- \HLvariable, set size --- \HLsetSize, target location --- \HLlocation, and layout --- \HLlayout)?  and 
(b) how robust are the visual variables with respect to set size (\HRsetSizegrid and \HRsetSizeNongrid) and target location (\HRdisToCengrid and \HRdisToCenNongrid)? 
To answer (b) we consider \textsc{Grid} and \textsc{Non-grid} separately because (i) the perceptual mechanisms involved in completing the task are likely different (e.g., length should be easier to perceive with alignment~\cite{Cleveland_1984,Heer2010}), (ii) these correspond to two distinct groups of visualizations (e.g., tables vs.\ scatterplots), and (iii) it simplifies the interpretation of the results.
The research questions translate into the following hypotheses:

\begin{description}[noitemsep]
    \item[\HLvariable] Different visual variables result in different task completion times. 
    For example, we expect that \textsc{Hue} results in shorter task completion times than \textsc{Shape}. 
    \textbf{Rationale:} Different visual variables rely on different perceptual mechanisms; empirical work in visualization is often about these differences~\cite{Cleveland_1984,Heer2010}. 
    Differences between visual variables are also likely for localization tasks.

    \item[\HLsetSize] Different set sizes result in different task completion times. 
    Specifically, the bigger the set size, the longer the task completion time. 
    \textbf{Rationale:} Much experimental work in visual search has found that an increase in the number of objects results in additional time (e.g.,~\cite{Francolini1979,Treisman_1980,Treisman_1982,Treisman1985,Treisman_1988, Treisman1985Preattenive,Moraglia1989,DZmura1991,Kristjnsson2001,Wolfe2021, Ware_2012}).
    
    \item[\HLlocation] Different target locations (\textsc{Center} or \textsc{Periphery}) result in different task completion times. 
    Specifically, centered targets result in shorter task completion times than peripheral targets. 
    \textbf{Rationale:} Targets in \textsc{Center} are likely in the fovea, which should reduce gaze movement and hence completion time.
    
    \item[\HLlayout] Different layouts result in different task completion times. 
    Specifically, \textsc{Grid} result in shorter task completion times than \textsc{Non-grid}. 
    \textbf{Rationale:} Previous empirical results show that grid layout supports magnitude comparison with some visual variables~\cite{Cleveland_1984,Heer2010}, which might also lead to faster times.

    \item[\HRsetSizegrid] Different visual variables result in different locatability robustness with respect to set size, in grid layouts.
    For example, we expect that \textsc{Hue} is more robust than \textsc{Shape}. 
    \textbf{Rationale:} Visual variables rely on different perceptual mechanisms (e.g., foveation), which likely result in different levels of robustness. 

    \item[\HRsetSizeNongrid] Different visual variables result in different locatability robustness with respect to set size, in non-grid layouts.
    For example, we expect that \textsc{Hue} is more robust than \textsc{Shape}. 
    \textbf{Rationale:} Same as in \HRsetSizegrid, but alignment is also known to affect visual variables such as \textsc{Length} and not others~\cite{Heer2010}.
    
    \item[\HRdisToCengrid] Different visual variables result in different locatability robustness with respect to the distance of the target to the center point of the display, in grid layouts. 
    For example, we expect that \textsc{Hue} is more robust than \textsc{Shape}.
    \textbf{Rationale:} Some changes in visual variables might be detectable from further away.

    \item[\HRdisToCenNongrid] Different visual variables result in different locatability robustness with respect to the distance of the target to the center point of the display, in non-grid layouts. 
    For example, we expect that \textsc{Hue} is more robust than \textsc{Shape}.
    \textbf{Rationale:} Same as in \HRdisToCengrid but with the possible effect of alignment.
\end{description}

\section{Experiment 1}
\label{sec:experiment1}
We test our hypotheses with two crowdsourced experiments. 
Experiment 1 is between-subjects, with half of the participants carrying out \textsc{Grid} trials and the other half \textsc{Non-grid} trials. 
We divide participants into two groups because sustaining participants' attention over long periods in crowdsourced research is challenging~\cite{Gritz2019}.

\subsection{Experiment 1 -- Methodology}

Here we describe the conditions, task, apparatus, measurements, analysis approach, and procedure for experiment 1.

\subsubsection{Conditions}
\label{sect:study_design:Conditions}

\begin{figure}[t!]
    \centering
    \includegraphics[width=1\linewidth]{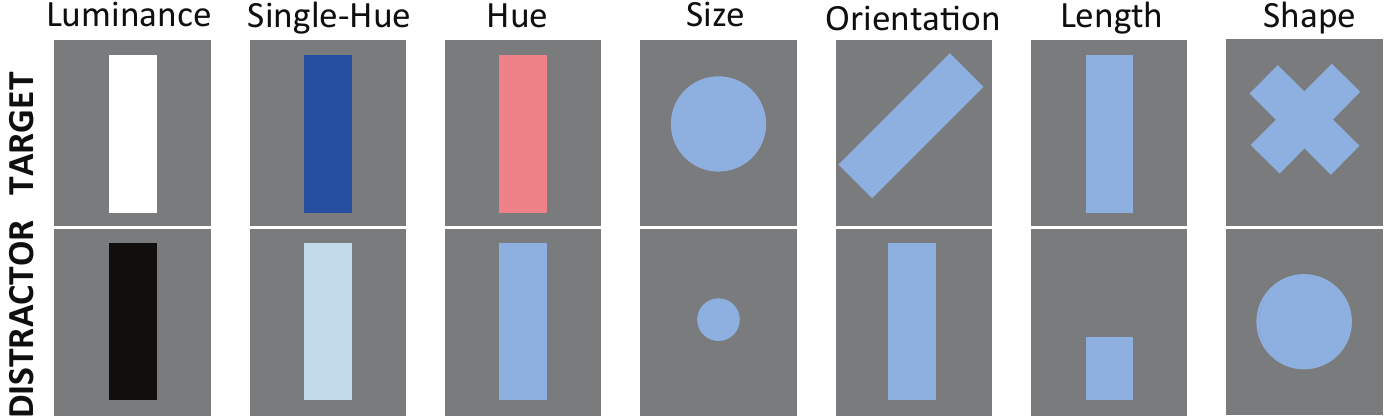}
    \caption{The target and distractor for each visual variable.}
    \label{fig:Features}
\end{figure}

\inlinesection{Visual variables}
had two levels each, based on three objectives: 
(i) to maximize the perceptual range of differences between levels; 
(ii) to maintain consistency with levels tested in previous work; and
(iii) to test variations that are practical in visualization. 
Some variables such as size or length can have levels with virtually infinite differences between them---e.g., a pixel-sized object vs.\ a screen-sized object for size---which is where practical considerations (iii) become important. Figure~\ref{fig:Features} shows the target and distractor we used for each visual variable.

For color-based variables, targets and distractors were rectangular shapes, subtending 0.5 degrees in height and 0.15 degrees in width.
For \textsc{Luminance}, targets were white (\texttt{\#ffffff}) and distractors black (\texttt{\#000000}). 
For \textsc{Single-Hue}, we used the second (\texttt{\#bdd7e7}, distractors) and the fifth (\texttt{\#08519c}, targets) colors from the D3 discrete (5) sequential Single-Hue color\footnote{\url{https://observablehq.com/@d3/color-schemes}}. 
For \textsc{Hue}, we used Healey et al.~\cite{Healey_1995}'s color values with red (\texttt{\#ee9590}) for targets and blue (\texttt{\#87b0e5}) for distractors. 
We used that same blue color for both targets and distractors when testing the other visual variables, for consistency.

For \textsc{Orientation} and \textsc{Length}, targets and distractors had the same rectangular shape we used for color-based variables.
For \textsc{Orientation}, targets were rotated 45 degrees clockwise. 
For \textsc{Length}, the length of the distractors was 0.4 times the length of the targets. 

For \textsc{Size}, we used circles instead of rectangles, with targets subtending 0.3 degrees and the area of the distractors being 0.2 times that of the targets. 
For \textsc{Shape}, distractors were the same circles used as targets for \textsc{Size}, and targets were crosses. 
To ensure uniformity in perceived size, the area of the crosses matched that of the circles.

\inlinesection{Set size}
had four levels: 12, 48, 192, and 768.

\inlinesection{Layout} 
had two levels: \textsc{Grid} and \textsc{Non-grid}. For \textsc{Grid}, all objects were equidistantly aligned vertically and horizontally. For \textsc{Non-grid}, the objects were randomly positioned uniformly over the plane. Both layouts prevent objects from overlapping.

\begin{figure}[t!]
    \centering
    \begin{tabular}{cc}
    \includegraphics[width=0.32\linewidth]{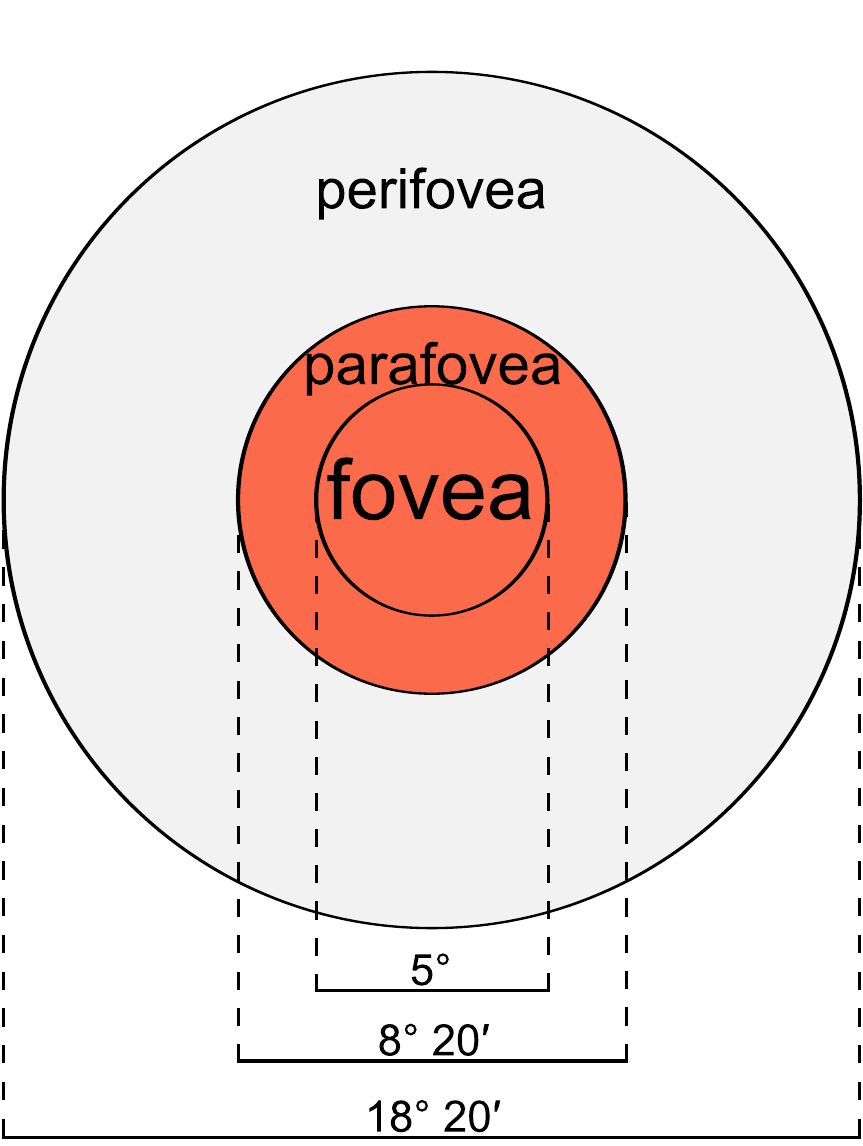} &
    \includegraphics[width=0.52\linewidth]{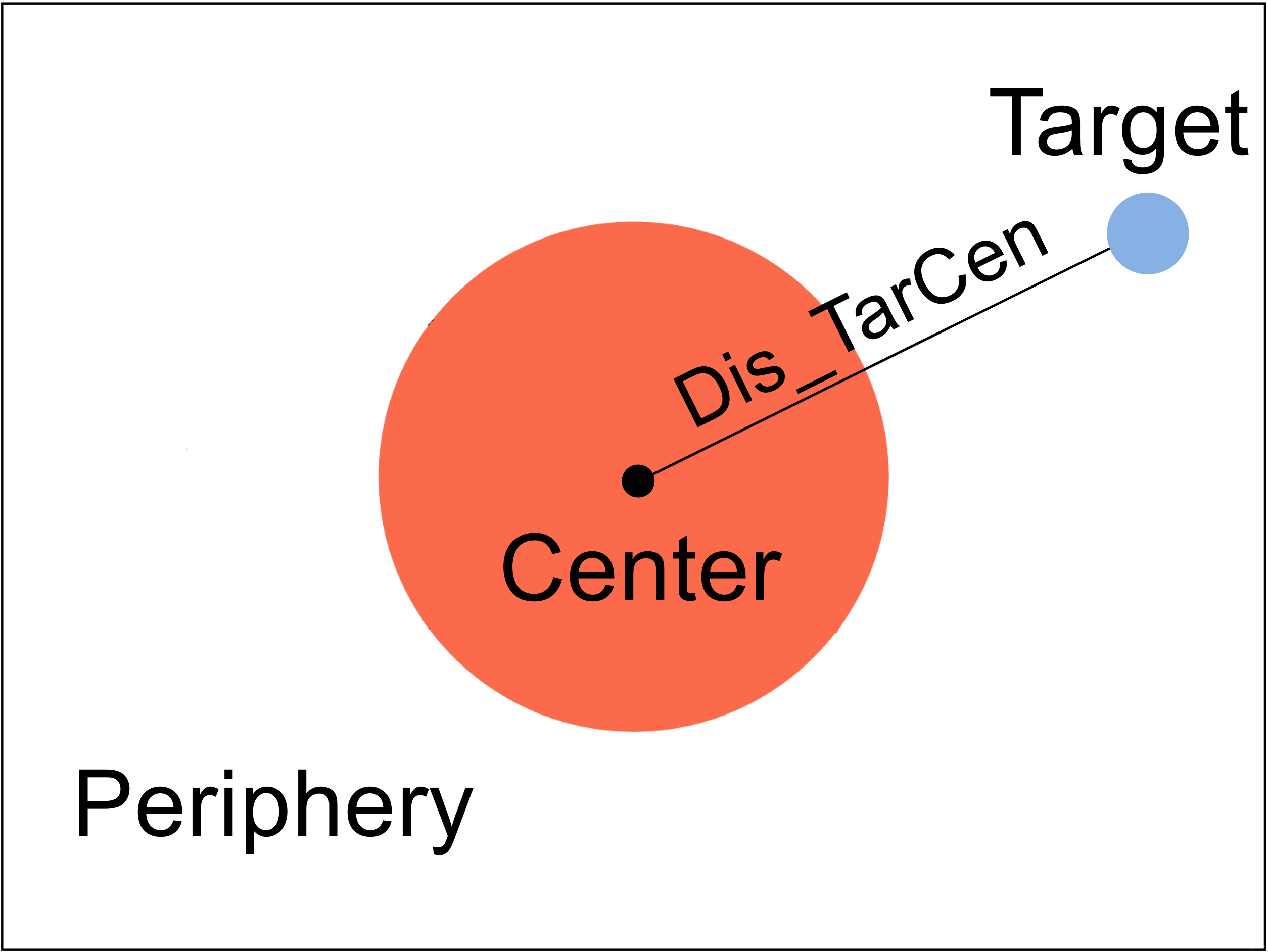}
    \end{tabular}
    \caption{Left: illustration of parafovea and fovea (orange), redrawn from ~\cite{wikimedia_Macula}. Right: area division for target location (orange: \textsc{Center}, white: \textsc{Periphery}).}
    \label{fig:areaDivision}
\end{figure}

\inlinesection{Target location} 
had two levels: \textsc{Center} and \textsc{Periphery}, with equal number of trials in each. 
Objects in \textsc{Center} were within the parafoveal+foveal regions of the visual field, while those in \textsc{Periphery} were outside these regions (see Figure~\ref{fig:areaDivision}). 
Participants did not know about this boundary.
For analysis purposes, we also calculated the geometric distance between the target and the center point of the display for each trial (Dis\_TarCen), which is a continuous variable.

\subsubsection{Task}
\begin{figure}[t!]
    \centering
    \includegraphics[width=\linewidth]{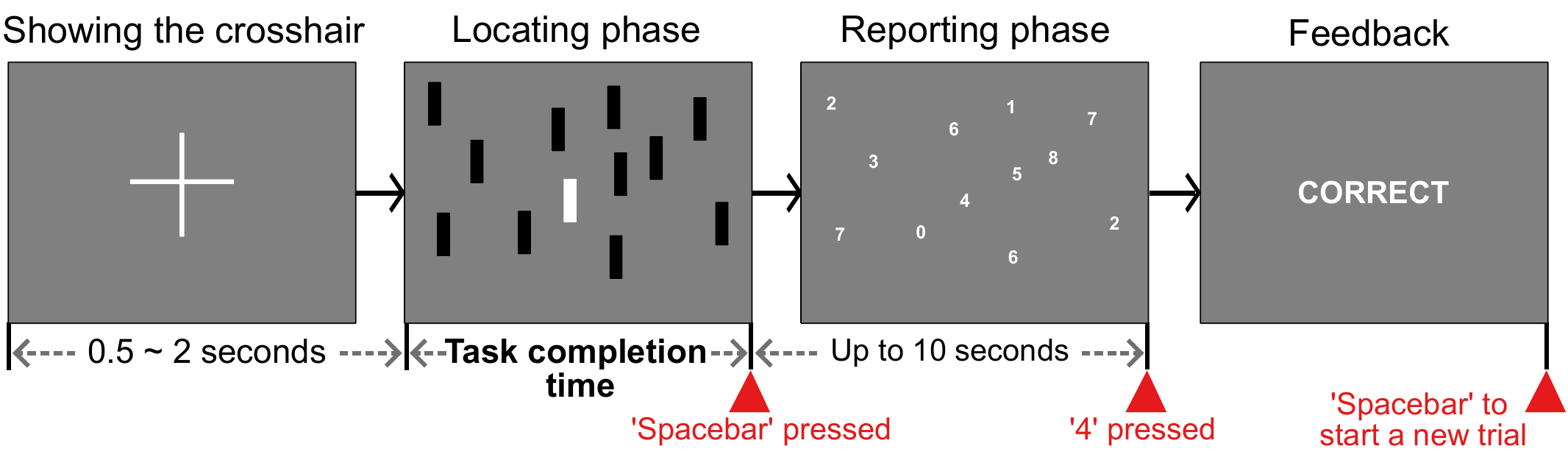}
    \caption{Flowchart of an example trial with \textsc{Luminance} in our study. 
    }
    \label{fig:procedure}
\end{figure}

There was a single localization task in the experiment, that included two phases: a \emph{locating phase}
and a \emph{reporting phase}. Figure~\ref{fig:procedure} shows an example of this task for \textsc{Luminance} in \textsc{Non-grid} layouts.

A trial begins showing a fixation crosshair for a random duration between 0.5 and 2 seconds (this prevents getting into a rhythmic response pattern across trials). 
Then, the stimulus screen appears and the timer for the locating phase of the trial starts. 
Each stimulus is an image with a neutral gray background that contains a multiplicity of identical objects and a target that is different in the level of one of its visual characteristics. 
Once they have located the target, the participant presses the space bar, which stops the timer for the trial.
The display then replaces the distractors and the target with digits (0--8) for the reporting phase. 
The participant has 10 seconds to enter the digit that appeared at the target’s location. 
The reporting phase ensures that the participant correctly located the target (inspired by~\cite{Vickery2005}) but the time spent in this phase is not included in the task completion time. 
Each trial ends with feedback indicating whether the response was correct.
\added{Providing trial feedback is common practice in visual search experiments (e.g., \cite{DZmura1991, Treisman1985, Healey_1995}); it helps maintain participants' attention and engagement, reduces frustration, and deincentivizes random responses.}

\subsubsection{Apparatus}

Participants redirected from Prolific reached a custom web-based experimental software built with Psychopy~\cite{Peirce2019} and deployed on Pavlovia\footnote{\url{https://pavlovia.org/docs/home/about}}. 
The application ran only on Microsoft Windows systems.

\subsubsection{Measurements and Models}
\label{sect:study_design:modes}
The main measurement is the time between stimulus onset and the participant signaling that they located the target---task completion time (CT). Since CT is typically long-tailed, we log-transform it for analysis (i.e., we analyze $LogCT = log(CT)$). The means and intervals reported henceforth are back-converted estimations in the logarithmic domain. 

We model noise as a Student-t distribution (similar to the Normal distribution, but robust to outliers~\cite[p.458]{kruschkeDoingBayesianData2014}). 
We use two main models to test our hypotheses. The first considers factors as nominal linear contributors to the mean ($\mu$) of the Student-t distribution (\autoref{model1}). 

\begin{equation}
\begin{split}
\label{model1}
\mu =  &a_0  + \overrightarrow{a_1}\cdot \overrightarrow{x_{Layout} }+ \overrightarrow{a_2}\cdot \overrightarrow{x_{TargetLocation}} + \overrightarrow{a_3}\cdot \overrightarrow{x_{SetSize}} \\&+  \overrightarrow{a_4}\cdot \overrightarrow{x_{VisualVariable}} + \overrightarrow{a_5}\cdot \overrightarrow{x_{Participant}} + interactions
\end{split}
\end{equation}

\changeColor{
\noindent The predicted mean ($\mu$) is a baseline ($a_0$) plus deflections ($\overrightarrow{a_1}$--$\overrightarrow{a_5}$) due to the level of the five factors (layout, target location, set size, visual variable, participant), and deflections due to all interactions between all factors excluding participant, which is a random factor.
Variance is estimated individually for each combination. 
We modified the model slightly because the pre-registered model was not appropriate for a between-subjects factor. The new model attributes systematic differences between groups to the layout factor~\cite[p.430]{kruschkeDoingBayesianData2014}. See the supplementary materials for details of the model modification.}

The second model considers either set size or Dis\_TarCen as continuous ($x_{Con}$) instead of nominal (\autoref{model2}).

\begin{equation}
\label{model2}
\mu =  a_0 + \overrightarrow{a_{VV}}\cdot \overrightarrow{x_{VV}} + \overrightarrow{a_{Con}}\cdot \overrightarrow{x_{VV}} \cdot x_{Con}
\end{equation}
\changeColor{
The predicted mean ($\mu$) is a baseline ($a_0$) plus the deflection ($\overrightarrow{a_{VV}}$) due to each visual variable ($\overrightarrow{x_{VV} }$), and plus the product of the slope ($\overrightarrow{a_{Con}} \cdot \overrightarrow{x_{VV} }$) and the continuous value ($x_{Con}$). $\overrightarrow{a_{Con}} \cdot \overrightarrow{x_{VV} }$ means that the slope is modulated by each visual variable ($\overrightarrow{x_{VV} }$).
The slopes of LogCT times the log-transformed set size and of LogCT times Dis\_TarCen are measures of robustness with respect to set size and target location.
}

\subsubsection{Analysis Approach}\label{Analysis-Approach}
We pre-registered the experiment before data collection.\footnote{\url{https://doi.org/10.17605/OSF.IO/YWU9D}} 
Numbering and naming conventions here differ from the pre-registration for clarity.   

We apply a Markov-Chain Monte Carlo (MCMC) Bayesian approach 
(see~\cite{kruschkeDoingBayesianData2014} and~\cite{gelmanBayesianDataAnalysis2014})
because it avoids some of the pitfalls of significance testing and p-values~\cite{cummingNewStatisticsWhy2014,klineSignificanceTestingStatistics2013} and because it offers more flexibility in modeling and test design. 
We chose uninformative priors, checked chain for good mixture, and ensured that equivalent sample sizes were $>10,000$ with \emph{psrf} between 0.97--1.04 for all parameters of interest. 
We consider a hypothesis strongly supported if its Bayesian probability is above 95\%, and weakly supported if it is above 90\%. 
Conversely, a probability $<5\%$ is a strong rejection and $<10\%$ is a weak rejection. 
We carried out the analysis with JAGS v.4.3.0~\cite{plummer2003jags} in R v.4.1.3 (2022-03-10)~\cite{RPackage} interfaced by the RJAGS library v. 4-13~\cite{rjags}. The analysis scripts are in the pre-registration and in the supplementary materials. 

\subsubsection{Procedure}
\label{sec:study_design:procedure}
Participants self-screened for typical or corrected acuity, typical color vision, and English proficiency. 
After consent, participants calibrated screen size and viewing distance through \emph{Virtual Chinrest}~\cite{Li2020}
to ensure that stimuli appeared with approximately the same visual angle for each participant.
They then read the instructions and completed four training trials. These differed from the experimental conditions to avoid bias: distractors were color-filled circles and targets had a donut shape.

Each participant carried out 28 blocks (4 set sizes $\times$ 7 visual variables), each with 8 trials (4 \textsc{Center}, 4 \textsc{Periphery}, in unpredictable order), in their assigned layout (224 trials in total).
Block order for each participant was counterbalanced with a Latin-square.
The screen showed which visual variable would be used before the start of each block. 
Incorrect trials and trials participants marked as mistakes (if they accidentally pressed space bar before they located the target) were re-trialed at the end of the block until correctly completed. 
After each block, participants could take a break. 
At the end of the experiment, they were thanked for their participation and redirected to Prolific.

\subsection{Experiment 1 -- Participants}
We recruited 112 participants from Prolific. 
To be eligible, participants had to have an approval rate $>95\%$, be proficient in English and have typical or corrected acuity and color vision. 
58 males and 54 females aged 19--68 years of age (mean~31.8, median~29) participated.
Participants received 6 GBP for their participation. All participant sessions were manually verified and approved by the research team.

\subsection{Experiment 1 -- Confirmatory Results}
\label{sec:results}
The experiment took 47 minutes on average (min 19, max 99). As expected, accuracy was nearly perfect: 98.1\% of trials were correct on first attempt and no condition fell below 96\%. 
As per the pre-registration, we report each hypothesis result by marginalizing over other factors to isolate the effect of one factor. For example, to determine the impact of layout on CT, we marginalize the effects of set size, target location, and visual variable.  
We employ the term \emph{partially supported} if only some of the differences involved in a hypothesis are supported.

\subsubsection{Locatability}

\begin{figure}[t!]
    \centering
    \includegraphics[width=\columnwidth]{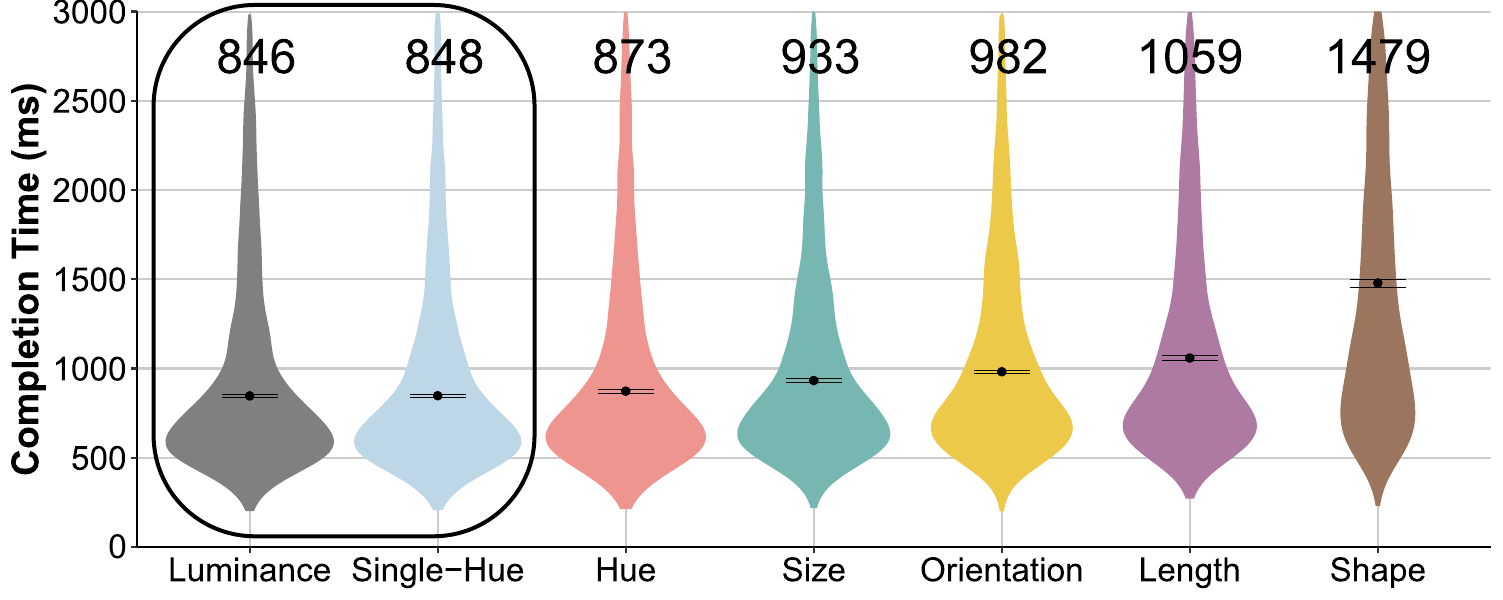}
    \caption{\textbf{Experiment 1}: Estimated Completion Time (black dot and floating numbers, in milliseconds) and 95\% High-Density Interval (HDI, whiskers) for each visual variable. Colored violins show data distribution (not estimated mean distribution). A rounded rectangle outline enclosing several conditions indicates that they are not statistically distinguishable. All other pairwise comparisons are distinguishable with probability $>95\%$.} 
    \label{fig:H4}
\end{figure}

\begin{figure}[t!]
    \centering
    \includegraphics[width=\columnwidth]{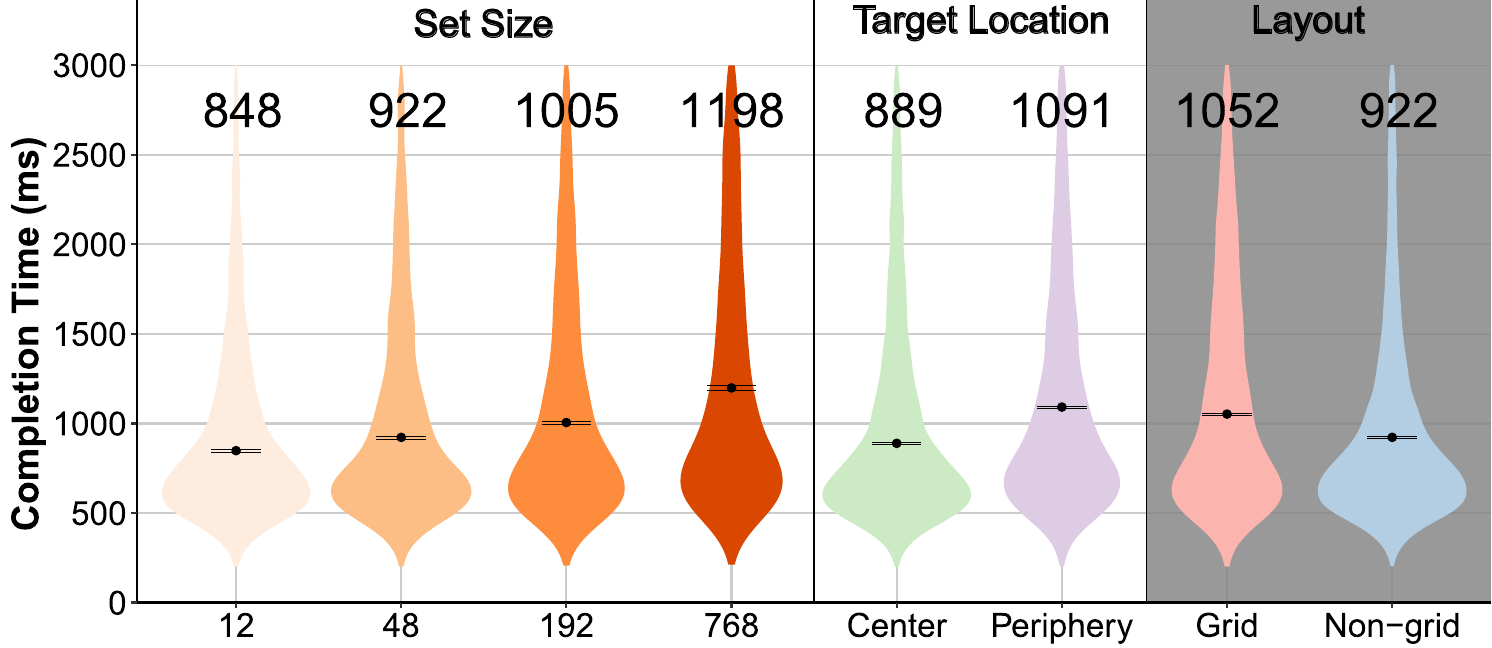}
    \caption{\textbf{Experiment 1}: Task CT for different set sizes (left), target locations (middle), and layouts (right) in experiment one. All pairwise comparisons $>95\%$ probability. Other notation as in Figure~\ref{fig:H4}}.
    \label{fig:H123}
    \vspace{-.75em}
\end{figure}

\inlinesection{\HLvariable}is strongly but partially supported, with strong evidence ($>95\%$ probability) of CT differences between all visual variables, except between \textsc{Luminance} and 	\textsc{Single-Hue} (Figure~\ref{fig:H4}).
\footnote{Violin plots show estimated means and intervals above their center of gravity. This is because means are estimated in the log-transformed domain and because the distributions are still somewhat long-tailed despite the transformation.}
The probability that CT are shorter with \textsc{Luminance} than with \textsc{Single-Hue} is 57.8\% (inconclusive), with an average difference of 2 ms. 

\inlinesection{\HLsetSize} is strongly supported (all pairwise comparisons show differences---Figure~\ref{fig:H123}-left), with longer CTs for larger set sizes. CT are on average 350 ms shorter with the 12 set size than with 768. 

\inlinesection{\HLlocation} is strongly supported, with CT 202 ms shorter for \textsc{Center} targets than with \textsc{Periphery} targets (Figure~\ref{fig:H123}-middle).

\inlinesection{\HLlayout} is strongly rejected, 
with CT 130 ms longer with \textsc{Grid} than with \textsc{Non-grid} (Figure~\ref{fig:H123}-right). However, this is likely confounded by between-subjects bias, as we discuss in Section~\ref{experiment1:discussion:layout}.

\subsubsection{Locatability Robustness}
By analyzing the slope of 
LogCT as $Log_{10}(set\ size)$  
and 
Dis\_TarCen vary, 
we find all locatability robustness hypotheses partially supported. 

\begin{figure}[t!]
    \centering
    \includegraphics[width=\columnwidth]{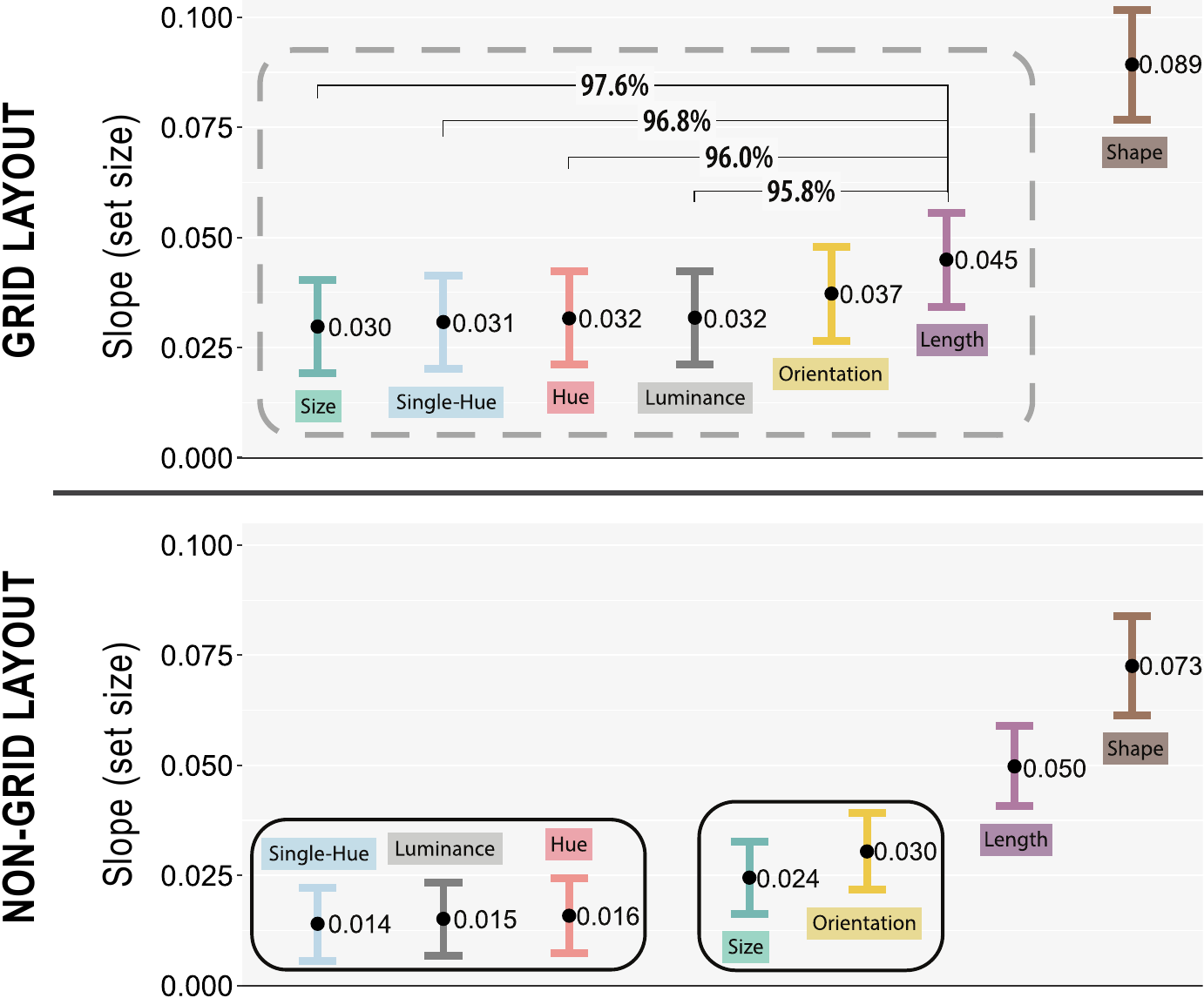}
    \caption{\textbf{Experiment 1}: Estimation of robustness (slope) of LogCT with respect to $Log_{10}(set\ size)$ (black dot, floating text) and 95\% HDIs, for each visual variable. 
    Dashed outlines indicate only some pairwise comparisons are statistically distinct. Solid outlines as in Figure~\ref{fig:H4}.}
    \label{fig:H56}
\end{figure}

\inlinesection{\HRsetSizegrid} is partially supported (see Figure~\ref{fig:H56}-top).
The results reveal two groups based on locatability robustness to set size, for grid layouts. 
Group 1 includes all variables but \textsc{Shape}, and within group 1, all variables but \textsc{Orientation} are more robust than \textsc{Length}.
Group 2 includes \textsc{Shape}, which is less robust than all other variables. 
    
\inlinesection{\HRsetSizeNongrid} is partially supported (see Figure~\ref{fig:H56}-bottom).
The results reveal four groups based on locatability robustness to set size, for non-grid layouts.
Group 1 includes the three color-based variables (\textsc{Single-Hue}, \textsc{Luminance}, and \textsc{Hue}).
Group 2 includes \textsc{Size} and \textsc{Orientation}, group 3 \textsc{Length}, and group 4 \textsc{Shape}.

\begin{figure}[t!]
    \centering
    \includegraphics[width=\columnwidth]{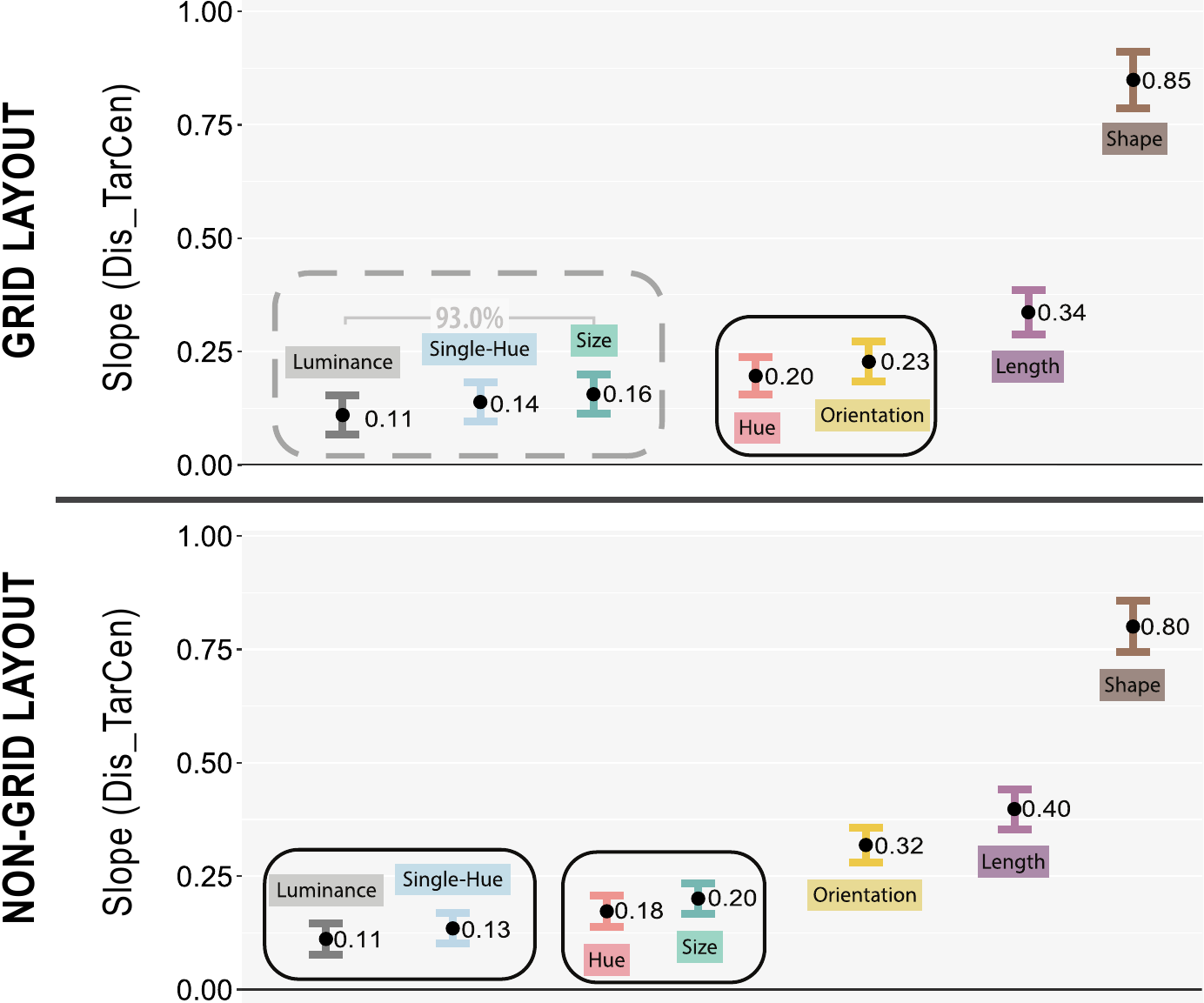}
    \caption{\textbf{Experiment 1}: Estimation of robustness (slope) of LogCT with respect to Dis\_TarCen, per visual variable. Other notation as in Figure~\ref{fig:H56}.}
    \label{fig:H78}
\end{figure}

\inlinesection{\HRdisToCengrid} is partially supported (see Figure~\ref{fig:H78}-top).
The results reveal four groups based on locatability robustness to distance between target and center, for grid layouts.
Group 1 includes \textsc{Luminance}, \textsc{Single-Hue}, and \textsc{Size}, and within group 1, \textsc{Luminance} is weakly more robust than \textsc{Size}. 
Group 2 includes \textsc{Hue} and \textsc{Orientation}, group 3 \textsc{Length}, and group 4 \textsc{Shape}. 
    
\inlinesection{\HRdisToCenNongrid} is partially supported (see Figure~\ref{fig:H78}-bottom). 
The results reveal five groups based on locatability robustness to distance between target and center, for non-grid layouts.
Group 1 includes \textsc{Luminance} and \textsc{Single-Hue}. 
Group 2 includes \textsc{Hue} and \textsc{Size}, group 3 \textsc{Orientation}, group 4 \textsc{Length}, and group 5 \textsc{Shape}.

\subsection{Experiment 1 -- Issues About Layout} \label{experiment1:discussion:layout}
We were surprised by the result of our test of \HLlayout (the overall comparison between \textsc{Grid} and \textsc{Non-grid}) because this went against our predictions. 
This would have been an interesting result, but we questioned its validity because we had found evidence of the opposite in pilot studies (participants performed better with grid layouts). 

In the search for an explanation of the incongruity, we realized that the sessions of participants assigned to non-grid layouts were all carried out before those of participants assigned to grid layouts. 
This could have introduced unwanted bias in the only factor that was between-subjects (e.g., weekend participants being inherently slower than weekday participants).
This led us to carry out Experiment 2.

\section{Experiment 2} \label{sec:experiment2}
Experiment 2 is a semi-replication of Experiment 1 addressing Section~\ref{experiment1:discussion:layout}. It enables: (i) testing \textsc{Grid} vs. \textsc{Non-grid} differences in a within-subjects design less vulnerable to bias, and (ii) verifying that the issue did not invalidate other tests. 
Experiment 2 was pre-registered.\footnote{\url{https://doi.org/10.17605/OSF.IO/JTS6D}}

\subsection{Experiment 2 -- Methodology}

Experiment 2 was similar to Experiment 1, except that each participant completed both \textsc{Grid} and \textsc{Non-grid} trials. 
The tradeoff is that Experiment 2 included only two set sizes (12 and 768) so that the overall duration was consistent with Experiment 1.
There were 28 blocks in total: 2 layouts $\times$ 2 set sizes $\times$ 7 visual variables, each block containing 8 trials (4 \textsc{Center}, 4 \textsc{Periphery}, in an unpredictable order). 

\subsection{Experiment 2 -- Participants}

We recruited 112 participants from Prolific with the same criteria and compensation as in Experiment 1. Participants of Experiment 1 were ineligible to participate. Due to a technical issue, the demographic information of one participant was not recorded. Participants (54 males, 57 females) were 18--72 years of age (mean~34.3, median~31).


\subsection{Experiment 2 -- Confirmatory Results}
The experiment took on average 45 minutes (min~19, max~92). The accuracy was again nearly perfect. Across all trials, 98.2\% were correctly answered on the first attempt, with no condition falling below 94\%.
The data analysis follows the same approach as in Experiment 1. However, this time, we do not need to modify the first model (see Section~\ref{sect:study_design:modes}) because all factors are within participants. 

The key result in Experiment 2 concerns layout: this time, as expected, \HLlayout is strongly supported, with completion times being on average 101 ms shorter with \textsc{Grid} than with \textsc{Non-grid}.

\begin{figure}[t!]
    \centering
    \includegraphics[width=\columnwidth]{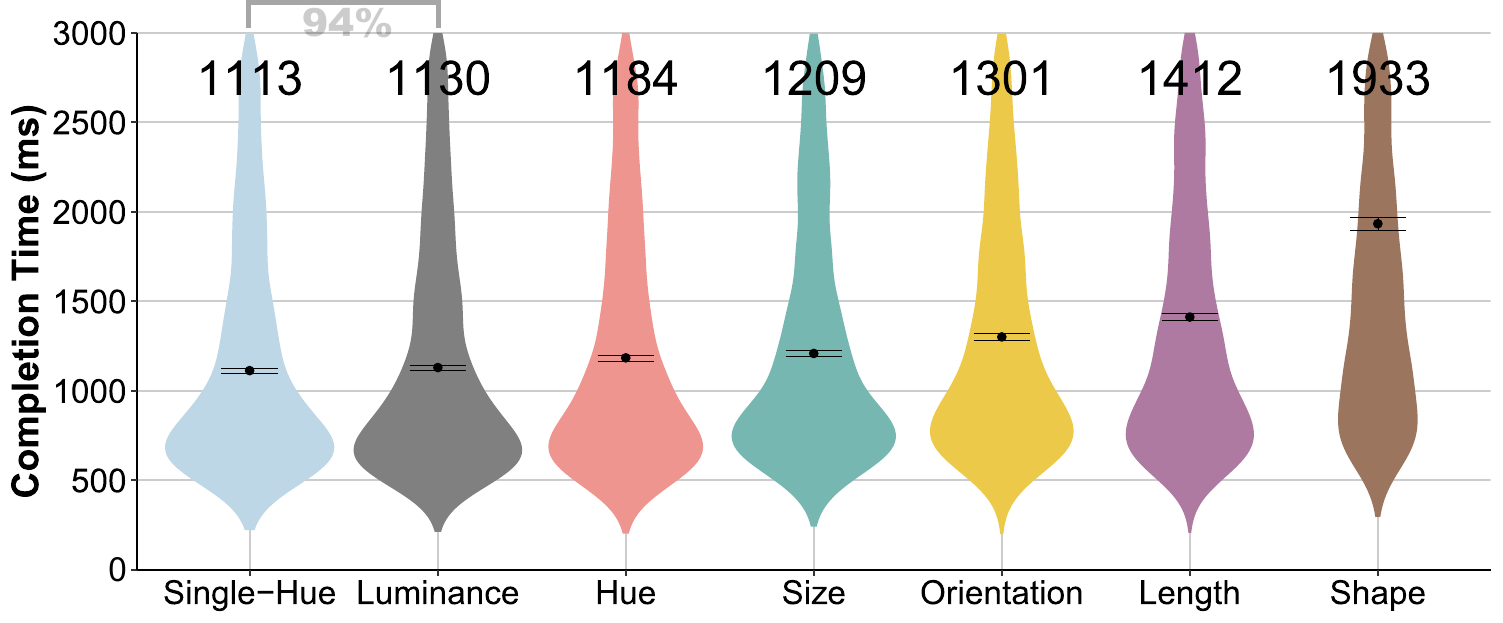}
    \caption{\textbf{Experiment 2}: Estimated average CT for the seven visual variables. 
    All pairwise comparisons have $>95\%$ probability except \textsc{Single-Hue} with \textsc{Luminance} (94\%). 
     Other notation as in Figure~\ref{fig:H4}.
    }
    \label{fig:H4S3}
\end{figure}

The other results are consistent with those of Experiment 1.
\HLvariable is again strongly supported, with strong evidence of differences in task completion times between all visual variables, except between \textsc{Luminance} and \textsc{Single-Hue}. 
Note that Experiment 2 does show a weak difference between \textsc{Luminance} and \textsc{Single-Hue} (see Figure~\ref{fig:H4S3}), with a 94\% probability that completion times are shorter for \textsc{Single-Hue} than for \textsc{Luminance} (average difference of 17~ms).

\begin{figure}[t!]
    \centering
    \includegraphics[width=\columnwidth]{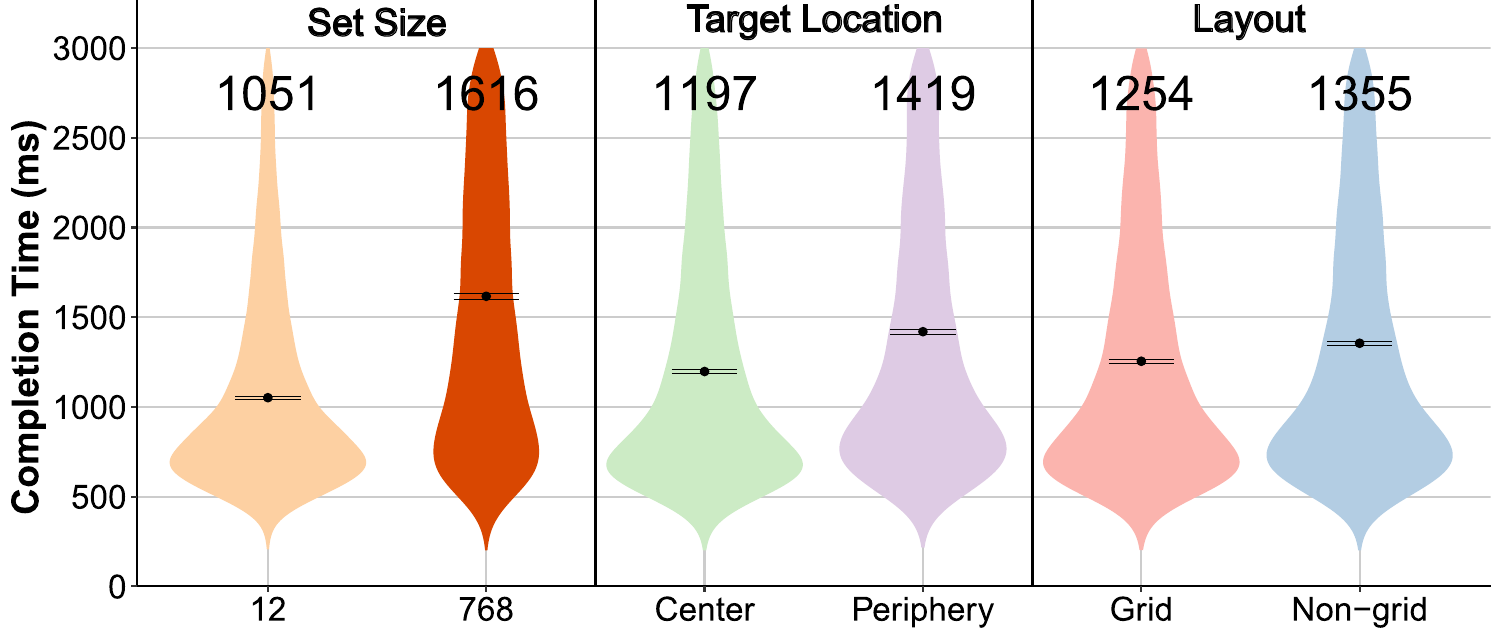}
    \caption{\textbf{Experiment 2}: Estimated average CT for different set sizes (left), target locations (middle), and layouts (right). All pairwise comparisons $>95\%$ probability. Other notation as in Figure~\ref{fig:H4}.}
    \label{fig:H123S3}
\end{figure}

The results for set size and target location (Figure~\ref{fig:H123S3}) are similar to Experiment 1's, with strong evidence (all $>95\%$ probability) of differences in CT for set size (\HLsetSize) and target location (\HLlocation).

We do not analyze locatability robustness with respect to set size because the independent variable (set size) has only two levels, which makes linear regression uninformative with just two data points. 
Results for locatability robustness to Dis\_TarCen are overall consistent with those of Experiment 1, with \HRdisToCengrid and \HRdisToCenNongrid still partially supported.

\begin{figure}[t!]
    \centering
    \includegraphics[width=\columnwidth]{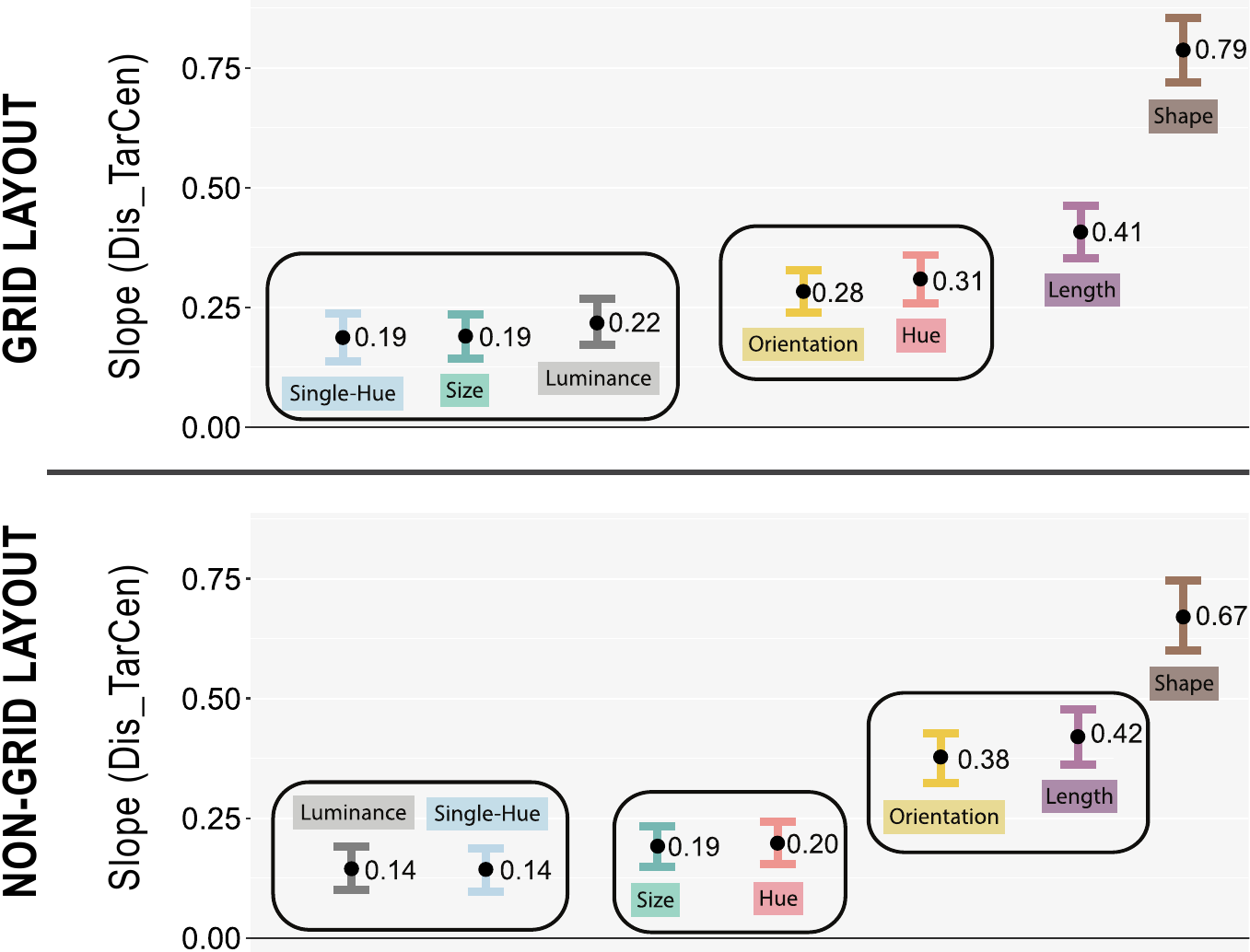}
    \caption{\textbf{Experiment 2}: Estimation of robustness (slope) of LogCT with respect to Dis\_TarCen, per visual variable. Other notation as in Figure~\ref{fig:H56}.}
    \label{fig:H78S3}
\end{figure}

For locatability robustness with \textsc{Grid} (Figure~\ref{fig:H78S3}-top), we find the same four groups as in Experiment 1 (Figure~\ref{fig:H78}-top), with only slight variations in the order within the first two groups.

For \textsc{Non-grid} (Figure~\ref{fig:H78S3}-bottom), the results are consistent with those of Experiment 1, except that there is no conclusive evidence distinguishing \textsc{Orientation} and \textsc{Length}.
Consequently, while the overall ordering of the visual variables' locatability robustness to Dis\_TarCen for \textsc{Non-grid} is consistent across experiments, Experiment 2 has four groups (instead of the five in Experiment 1). 

\subsection{Experiment 2 -- Exploratory Results on Layout}
\label{sec:exploratory-results}



Increased confidence in layout comparison validity allowed us to further explore how visual variables are affected by layout. The following results are not pre-registered and should be cautiously interpreted.

\begin{figure}[t!]
    \centering
    \includegraphics[width=1.0\linewidth]{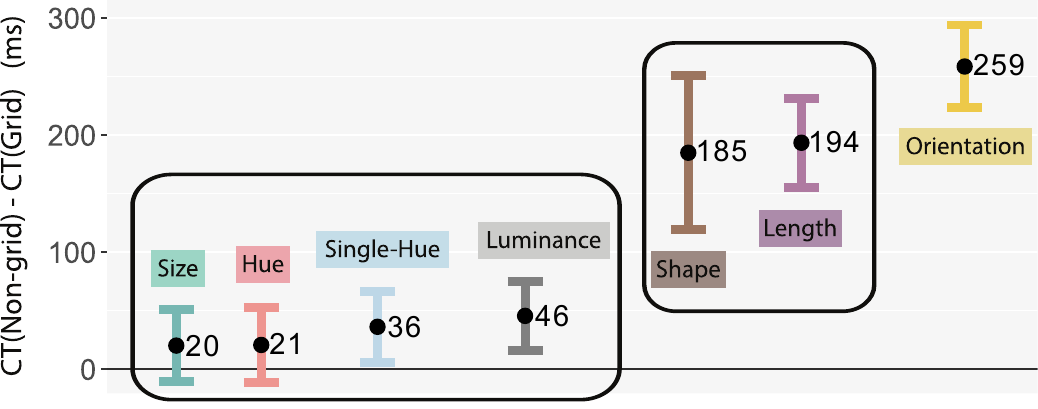}\\
    \caption{\textbf{Experiment 2}: Difference between \textsc{Grid} and \textsc{Non-grid} robustness with respect to set size.
    Other notation as in Figure~\ref{fig:H56}.}
    \label{fig:layoutDiff}
\end{figure}

Section~\ref{sec:locatability} describes how locatability robustness is not limited to continuous factors. 
Nominal factors such as the type of layout might affect the robustness of visual variables as well.
Here we consider whether variables are more or less robust with respect to layout. 
Since we only have two categories for layout, the degree to which locatability changes when layout changes is the difference in task completion times between \textsc{Non-grid} and \textsc{Grid} 
(see Figure~\ref{fig:layoutDiff}).
The closer the difference is to zero, the more robust a visual variable is to layout. 

The results show that all visual variables are negatively affected by non-grid layouts but to different extents (we identified three groups).
Group 1 includes \textsc{Size}, \textsc{Hue}, \textsc{Single-Hue}, and \textsc{Luminance} (CT with grid layouts shorter than with non-grid layouts by less than 50~ms on average).
Group 2 includes \textsc{Shape} and \textsc{Length} (CT with grid layouts shorter by 185~ms and 194~ms). 
Group 3 includes \textsc{Orientation} (CT with grid layouts shorter by 259~ms).

Additionally, visual variable ordering based on percentage change relative to grid layouts $(CT(\textsc{Non-grid}) - CT(\textsc{Grid}))/ CT (\textsc{Grid})$ is consistent with their robustness level with respect to layout: \textsc{Size}---1.7\%, \textsc{Hue}---1.7\%, \textsc{Single-Hue}---3.3\%, \textsc{Luminance}---4.1\%, \textsc{Shape}---10\%, \textsc{Length}---14.7\%, and \textsc{Orientation}---22\%.
This confirms that layout affects visual variables to varying degrees, both in absolute time change and in relative percentage of change.

\section{Discussion }
\label{sec:discussion}

\changeColor{
The aim of this work is to empirically contribute to the characterization of the effectiveness of visual variables, so that researchers and designers have a deeper understanding of visual variables for localization tasks. This should eventually lead to more effective visualization design. 
Notwithstanding the applied nature of this goal, it is crucial to leverage the hard-won knowledge from the extensive perceptual literature, despite that work in psychology tends to focus on discovering the principles of how the human perceptual system works rather than on their application to visual displays. 
Simultaneously, our work cannot only consider specific scenarios of applicability (e.g., recommendations for a kind of scatterplot) if it is to be generalizable and provide a useful foundation for the visualization community. 
Hence, we see our contributions as a bridge between perception research and its application in visualization. We seek both relevance to visualization researchers and generalizability rather than focusing on specific recommendations for design. More empirical work is necessary to systematically derive recommendations for practitioners from our findings.
}

\changeColor{
Consequently, we first discuss in Section~\ref{subsec:takesaways} the main takeaways of the experiments (in \textbf{boldface}), which focus on what the results mean for our understanding of visual variables. Only then we demonstrate, with a few examples, how these findings can have an impact on concrete visualization design situations (Section~\ref{subsec:implications}). 
}

\subsection{Locatability of Visual Variables: Lessons Learned}
\label{subsec:takesaways}

\changeColor{
The most general finding is that \textbf{different visual variables exhibit different locatability}. 
Color-based variables (\textsc{Luminance}, \textsc{Single-Hue}, and \textsc{Hue}) result in faster times for locating objects than \textsc{Size}, then \textsc{Orientation}, then \textsc{Length} and \textsc{Shape} (slowest). 
The advantage of color variables is consistent with prior research in visual search tasks~\cite{DZmura1991, Nagy1990, Healey1999} but the overall ranking from our data differs from previous rankings of the preattentiveness of variables~\cite{Haroz2012}\cite[p. 24]{TheHandbookofAttention}. 
}

\changeColor{
To our knowledge, ours is the first empirically derived ranking of the locatability of visual variables, analogous to the popular rankings of visual variables for difference estimation that are commonly included in visualization textbooks and courses (e.g.,~\cite{Cleveland_1984,Heer2010}\cite[p.102]{Munzner_2014}). 
This expands the empirical knowledge of how visual variables support localization, a necessary and common sub-task in many displays.
}

\changeColor{
We also found that \textbf{none of the visual variables tested exhibit perfect locatability robustness with respect to set size}: all slopes are larger than zero. 
This challenges the common belief that some variables are ``undoubtedly preattentive''~\cite{Haroz2012} or have a ``near zero slope''~\cite{Wolfe2021}. 
Visual variables are neither locatable or unlocatable, nor robust or not robust---echoing Wolfe's nuanced description of performance in visual search tasks~\cite{Wolfe1998Visual}. 
Most previous visual search studies tested relatively small numbers of distractors---to our knowledge, the largest set size examined in previous work is 54~\cite{Nagy1990}. 
The combination of testing small set sizes in previous experiments and the fact that time grows less-than-linearly with set size likely caused the misconceptualization of variables as dichotomous in \emph{efficiency} or as being driven by perceptual/cognitive processes supposedly immune to scale in the number of objects (preattentive processing). 
Nevertheless, knowing that localization is not perfectly robust to set size, and being able to estimate the time cost differences for common visual variables can make a difference in visualization, where the number of objects in a display is often counted in hundreds or thousands. Additionally, our measurements provide empirical support for the common choice of color over other variables for highlighting.
}

\changeColor{
Target location is the tested factor with the strongest effect: \textbf{for all variables we tested, the further from the current fixation point a target is, the longer it takes to locate it.} 
This might seem obvious if we assume that locating an object requires fixating the gaze on the position of the object, which takes longer the further it is from the current gaze location~\cite{Miniotas2000}.
However, we have not found previous empirical evidence of this result for localization (a study considering a similar problem with magnitude estimation task is in~\cite{Bezerianos2012}). 
Furthermore, \textbf{different visual variables have different robustness with respect to target location}, and the robustness ranking varies based on layout. 
In non-grid layouts, the robustness with respect to target location mirrors locatability, unlike in grid layouts, where \textsc{Size} is one of the most robust variables, overtaking \textsc{Hue} in the ranking. 
We think these deviations are due to the fundamental change in the task that uniform layouts enable: finding a larger object in a grid becomes akin to detecting a misalignment rather than detecting size variation or ink density.
}

\changeColor{
We also found in Experiment 2 that \textbf{locatability is better with grid layouts}. 
In our experience, layout is rarely a design option that is compared explicitly when choosing visualizations; in other words, it is rare to examine the design choice between a non-gridded mapping such as a scatterplot and a gridded mapping such as a table-based scalar field~\cite{Manteau2017Reading}. This is probably because other considerations, such as the form and resolution of the data, take precedence. 
However, the advantage of grid layouts might shed light on why tables~\cite{jieffect} and other gridded alternatives (e.g.,~\cite{lewisAchillesHeelScatter2023,park2023gatherplot,Keim2004,Langton2019}) are still receiving attention. 
This highlights a tradeoff between the general recommendation to use position as the most powerful variable to represent continuous attributes (due to magnitude estimation advantages~\cite{Cleveland_1984, Munzner_2014}), and the advantages of having a predictable grid (improved localization, as our results show). 
We suspect that, in many cases, advantages in locatability might override the accuracy benefits of using directly-mapped space. 
}

\subsection{Examples of Potential Impact on Design}
\label{subsec:implications}
\changeColor{
Although the general interpretation of the results in the previous section does not provide sufficient grounding for us to offer a general set of design recommendations, we demonstrate how our findings can have practical applicability with a few examples.
}

\begin{figure}[t!]
    \centering
    \includegraphics[width=\linewidth]{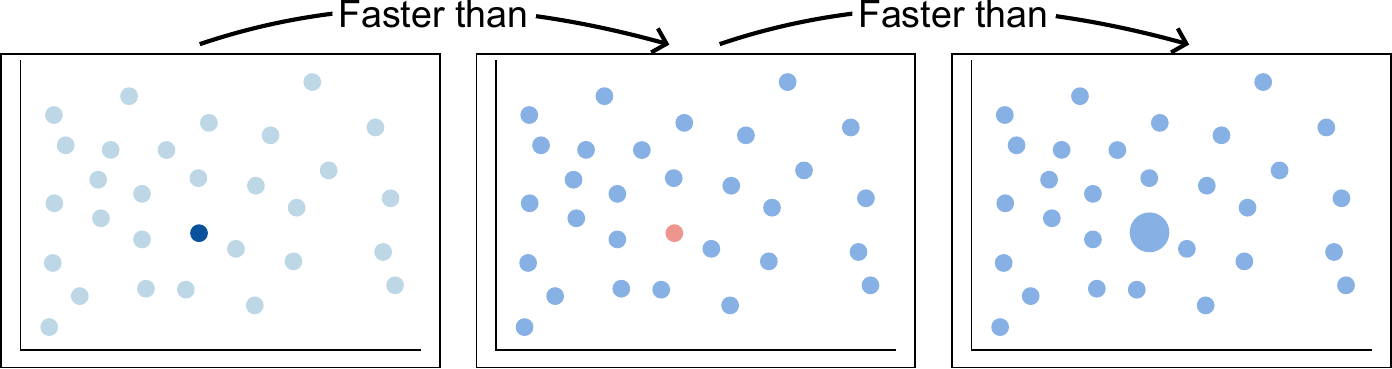}
    \caption{\textsc{Single-Hue} enables faster target localization than \textsc{Hue}, which in turn outperforms \textsc{Size} in a scatterplot.
    }
    \label{fig:exvis1}
\end{figure}

\changeColor{
Both experiments show the same general ranking of localization: locating with \textsc{Luminance}, \textsc{Single-Hue}, \textsc{Hue} and \textsc{Size} is faster than with \textsc{Orientation}, \textsc{Length} and \textsc{Shape}. 
If we were designing a scatterplot where we can select which variable to use for highlighting, our data indicates that using single-hue or luminance is best. 
If those are unavailable (luminance is problematic when the background is completely dark or light), hue is the next best option, then size (see Figure~\ref{fig:exvis1}). 
Any of the other variables will cause a noticeable slowdown, and using shape would cost almost twice the time on average.
}

\begin{figure}[t!]
    \centering
    \includegraphics[width=\linewidth]{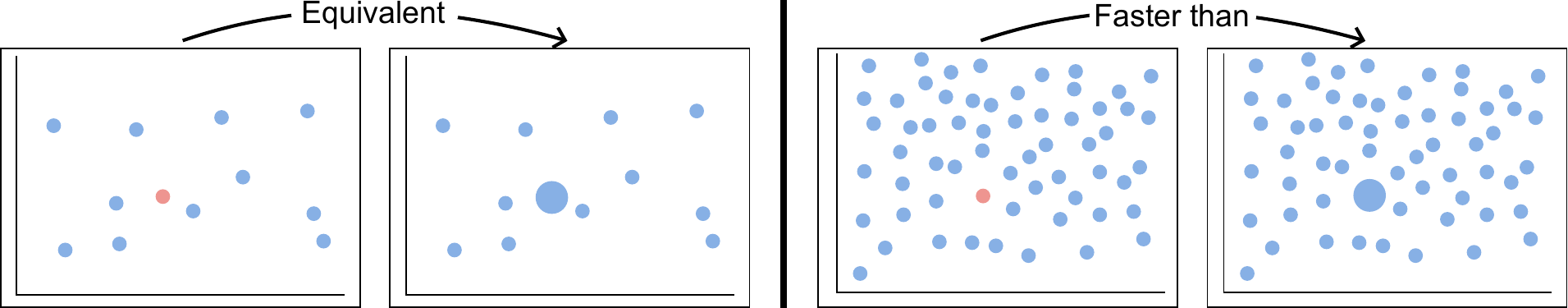}
    \caption{\textsc{Hue} is faster than \textsc{Size} with many objects (right), although they are similarly effective with small set sizes.}
    \label{fig:exvis2}
\end{figure}

\begin{figure}[t!]
    \centering
    \includegraphics[width=\linewidth]{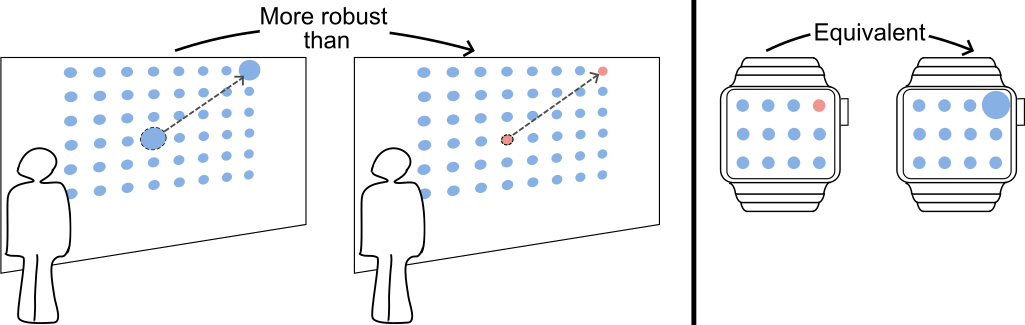}
    \caption{\textsc{Size} is more robust than \textsc{Hue} with respect to target location in gridded layout (left), but they are similarly effective on small screens where objects are all in the fovea and the number of objects is limited.}
    \label{fig:exvis4}
\end{figure}

\changeColor{
Since \textsc{Hue} is more robust than \textsc{Size} with respect to set size in non-grid layouts, \textsc{Hue} is better than \textsc{Size} for supporting localization in dense scatterplots (see Figure~\ref{fig:exvis2}-right).
However, our results also show that for displays with few objects, \textsc{Size} and \textsc{Hue} yield similar performances (Figure~\ref{fig:exvis2}-left). 
Still, when the display shows a grid layout, \textsc{Size} is more robust than \textsc{Hue} with respect to location, and therefore might be a better choice. 
If peripheral locations are common (e.g., in large displays) and it is important to not advantage locating foveal objects over peripheral objects, size is a better choice. 
But, if the highlighted objects are always in the fovea (e.g., with small displays like smartwatches), then both variables are equivalent (see Figure~\ref{fig:exvis4}).
}

\changeColor{
We provide a web-based interface (\url{https://locatability.github.io}) to access the quantitative model from our results. This model allows designers to compare specific situations such as the ones described above, for all of the tested conditions.
}

\subsection{Terminology and Tasks}\label{sec:diss:term_and_tasks}

It is legitimate to contemplate whether we need the newly defined terms to describe and characterize the localization task. 
We do not presume that locatability and locatability robustness invalidate or alter the meaning of the extensive body of previous work. 
\added{Nevertheless, we found it impractical to express our results with precision and clarity using existing terms.} 
For example, one could try to use \emph{search efficiency} instead of \emph{locatability robustness with respect to set size}, but that assumes that ``efficiency'' is always with respect to set size (and not, e.g., alignment, position, or other). 
\added{Similarly, the term preattentiveness presupposes that certain visual variables are inherently ``preattentive,'' an assumption that our results challenge. We therefore advocate for the use of locatability and locatability robustness in future research when (i) localization performance is the focus, (ii) factors beyond set size need to be considered in design decisions, and (iii) one seeks to avoid theoretical assumptions embedded in terms like preattentiveness. 
We believe that adopting these terms can enhance conceptual clarity and foster theoretical continuity in building a broader empirical foundation for visualization research.}

A related question is whether the localization task is at the right level of granularity to be useful. 
We think that localization, even though it encompasses the detection task\footnote{We consider that localization potentially includes the subtask of detecting whether there is a distinct target, foveating it, and then extracting information.} 
is more common in visualization than plain detection (e.g., a visual comparison between two points in a scatterplot will commonly require the foveation of both points in sequence). 
Importantly, the processes necessary for detection and foveation might happen in parallel and to different degrees of parallelism depending on the visual variable. 
Therefore, we cannot simply add the detection time to the foveation time for an accurate estimation of the localization time. Whether these subtasks are sequential or concurrent is an important question for future empirical research.

\section{Limitations and Future Work}
We have characterized the locatability and locatability robustness of seven visual variables according to three factors. 
To make this investigation feasible and focused, we controlled and simplified certain aspects of the study design. Accounting for all possible variables and factors in a single study would be impractical. For example, we prevented object overlap, even though overplotting may occur in real-world visualizations.
Fully characterizing human performance with visual variables requires further research, including investigating other variables such as texture, motion, and flicker that have been extensively studied in psychophysics~\cite{Driver1992,brown1965flicker} and that appear in visualizations~\cite{Healey1999,Huber2005, Heer2007, Chevalier2016, Esmaeili2023,Waldner2014,Gutwin_2017}. 
This also includes studying additional factors such as the number of targets,
the proximity of targets to distractors,
the density of distractors around a target,
and the degree of difference between targets and distractors.
\added{Additionally, investigating even larger set sizes (i.e., in the thousands) would help generalize our findings to visualizations with extreme numbers of visual objects.}

Our results also show an interesting relationship between length and size that is worthy of future study. 
In difference estimation tasks, length is generally ranked above size (area), but for localization, size consistently outranks length, in terms of both locatability and locatability robustness. 
This could be an artifact of our study because for \textsc{Length} and \textsc{Size} the choice of visual difference between target and distractors is subject to tradeoffs (e.g., too large circular targets for size would be impractical, and too long targets for length would limit the ability to display hundreds of elements in a space-constrained grid layout). 
Future evaluations might instead measure cross-variable comparisons such as the area ratio that corresponds to the same locatability (robustness) as in a given length ratio.

Methodologically, while we used Virtual Chinrest~\cite{Li2020} to control viewing conditions, the crowdsourced nature of the experiments means that we could not perfectly ensure that participants maintained a fixed viewing distance throughout. 
In-person studies can offer more control and incorporate additional measures such as gaze-position. Nonetheless, the large and diverse participant pool likely contributes to the reliability and generalizability of our findings.

\added{
Finally, although our results are likely to have broad application in visualizations where locating an object among many distributed in space is important---including scatterplots, small multiples, tables, and multi-row bar charts---the results will need to be replicated for variants that depart significantly in visual appearance from our stimuli (e.g., scatterplots with distributions that are highly non-uniform).
}
\section{Conclusions}
\label{Sect10}

We investigated human performance in localization tasks using different visual variables and considering how factors such as target location influence this performance. 
Through two experiments, we characterized the locatability and locatability robustness of seven visual variables---\textsc{Luminance}, \textsc{Single-Hue}, \textsc{Hue}, \textsc{Size}, \textsc{Orientation}, \textsc{Length}, and \textsc{Shape}---across three factors: set size, target location, and layout. 

We found that the visual variables we tested have different degrees of locatability, and that the three factors all have an impact on their locatability.
Although more empirical research is needed to establish a comprehensive characterization of visual variables for the localization task and corresponding, generalizable design guidelines, our results complement existing rankings of visual variables and provide a foundation that researchers and designers can build upon.



\acknowledgments{This project was funded by Canada’s NSERC (award 2019-05422 and 2020-04401). }

\bibliographystyle{PaciVis/abbrv-doi-hyperref-narrow}

\bibliography{PaciVis/bib-name-abbr, PaciVis/locatability}

\end{document}